\documentclass[twocolumn,prl,groupedaddress,showpacs,amsfonts,amssymb,amsmath,letterpaper]{revtex4}
\usepackage{amsmath}
\usepackage{graphicx}
\usepackage{amssymb}

\makeatletter

\begin{document}

\title{
Cavity-free Photon Blockade Induced by Many-body Bound States
}

\author{Huaixiu Zheng}
%\affiliation{\textit{Department of Physics, Duke University, P. O. Box 90305,
%Durham, North Carolina 27708, USA}}
%\affiliation{\textit{Center for Theoretical and Mathematical Sciences, Duke University,
%Durham, North Carolina 27708, USA}}
\author{Daniel J. Gauthier}
%\affiliation{\textit{Department of Physics, Duke University, P. O. Box 90305,
%Durham, North Carolina 27708, USA}}
\author{Harold U. Baranger}
\email{baranger@phy.duke.edu}
\affiliation{\textit{Department of Physics, Duke University, P. O. Box 90305,
Durham, North Carolina 27708, USA}}
%\affiliation{\textit{Center for Theoretical and Mathematical Sciences, Duke University,
%Durham, North Carolina 27708, USA}}

\date{\today}
 
\begin{abstract}
The manipulation of individual, mobile quanta is a key goal of quantum communication; to achieve this, nonlinear phenomena in open systems can play     a critical role. We show theoretically that a variety of strong quantum nonlinear phenomena occur in a completely open one-dimensional waveguide coupled to an $N$-type four-level system. We focus on photon blockade and the creation of single photon states in the absence of a cavity.  Many-body bound states appear due to the strong photon-photon correlation mediated by the four-level system. These bound states cause photon blockade, which can generate a sub-Poissonian single-photon source.
\end{abstract}


\maketitle

The exchange and control of mobile qubits of information is a key part of both quantum communication and quantum information processing. A ``quantum network'' is an emerging paradigm \cite{KimbleNat08,DuanRMP10} combining these two areas: local quantum nodes of computing or end-users linked together by conduits of flying qubits. The determinisitic approach to the interaction between the local nodes and the conduits relies on cavities to provide the necessary strong coupling. Indeed, strong coupling between light and matter has been demonstrated using cavities in both the classic cavity quantum electrodynamics (QED) systems \cite{MabuchiSci02} and the more recent circuit-QED implementations \cite{SchoelkopfNat08}. This has enabled the observation of nonlinear optical phenomena at the single photon level, such as electromagnetically induced transparency (EIT) \cite{MuckeNat10,AbdumalikovPRL10} and photon blockade \cite{BirnbaumNat05, LangPRL11}. Experiments have also demonstrated the efficient exchange of information between a stationary qubit (atom) and flying qubits (photons) \cite{BoozerPRL07,HofheinzNat09}. 
%However, cavities systems have other disadvantages, such as the complexity of stabilizing a cavity resonance to an atomic resonance. 
%However, scaling cavity systems to a multi-node quantum network is still technologically challenging because of the difficulties associated with connecting cavities and managing the losses.
However, scaling cavity systems to a multi-node quantum network is still challenging because of the difficulty of connecting cavities and managing losses.

A new scheme for achieving strong coupling between light and atoms (or artificial atoms) has been recently proposed based on one-dimensional (1D) \emph{waveguides} \cite{ChangPRL06,ShenPRL07,ChangNatPhy07,ZhengPRA10,RoyPRL11,KolchinPRL11}, dubbed ``waveguide QED'' \cite{ZhengPRA10,KolchinPRL11}.  
Tremendous experimental progress in achieving strong coupling has occurred in a wide variety of such systems: a metallic nanowire coupled to a quantum dot \cite{AkimovNat07},
%cold atoms trapped in a hollow fiber \cite{BajcsyPRL09}, 
a diamond nanowire coupled to a nitrogen-vacancy center \cite{BabinecNatNanotech10}, 
a photonic nanowire with an embedded quantum dot \cite{ClaudonNatPhoton10}, 
and a 1D superconducting transmission line coupled to a flux qubit \cite{AstafievSci10}.
% DAN SENTENCE: 
% Because of the tight confinement of optical fields in the transverse direction, strong coupling is achieved in these systems by ensuring that the majority of the spontaneously emitted light is guided into waveguide modes. 
% HAROLD SENTENCE:
In these systems, ``strong coupling'' means that the majority of the spontaneously emitted light is guided into waveguide modes; it is achieved through the tight confinement of optical fields in the transverse direction.
Furthermore, waveguide systems are naturally scalable for quantum networking \cite{KimbleNat08}. 
The key physical element introduced by the waveguide QED geometry is that the atom couples to a \emph{continuum} of modes. This relaxes the restriction of 
working with a narrow cavity bandwidth; more importantly, interaction with a continuum brings in novel many-body effects that have no analogue in a cavity.
%, notably the possibility of many-body bound states \cite{ShenPRL07, ZhengPRA10} when there are multiple photons.

%Here, in contrast, we investigate strong coupling deterministic effects that occur in the absence of a cavity, 
%specifically for a four-level system (4LS) coupled to a one-dimensional waveguide. 
%The variety of quantum  nonlinear phenomena that occurs---we focus on photon blockade, electromagnetically induced transparency, creation of single photon states, and strong bunching and anti-bunching---suggests that cavity-free deterministic quantum networks can be pursued. Indeed, recent leaps in experimental capabilities in several areas \cite{AkimovNat07,ClaudonNatPhoton10,AstafievSci10,BajcsyPRL09,BabinecNatNanotech10} mean that such phenomena could be observed in systems which are presently available. 

In this work, we show that the nonlinear optical phenomena EIT, photon blockade, and photon-induced tunneling emerge in a waveguide system for parameters \cite{AstafievSci10} that are currently accessible.
% \cite{AkimovNat07,ClaudonNatPhoton10,AstafievSci10,%BajcsyPRL09,
% BabinecNatNanotech10}. 
For these dramatic and potentially useful nonlinear effects, it is necessary to consider a four-level system (4LS) rather than simply a two- or three-level system.
Photon blockade and photon-induced tunneling have a completely different origin here from the cavity case \cite{BirnbaumNat05, FaraonNatPhy08}: they are produced by \emph{many-body bound states} \cite{ShenPRL07,ZhengPRA10,RoyPRL11,NishinoPRL09}, whose amplitude decays exponentially as a function of the relative coordinates of the photons.
Such states do not exist in cavities because a continuum of modes in momentum space is needed for the formation of bound-states in real space. We demonstrate the capability of such a system to generate a single-photon source, which is crucial for quantum cryptography and distributed quantum networking. Our work thus opens a new avenue toward the coherent control of light at the single-photon level based on a cavity-free scheme.

 \begin{figure}[t]
 \centering
 \includegraphics[width=0.48\textwidth]{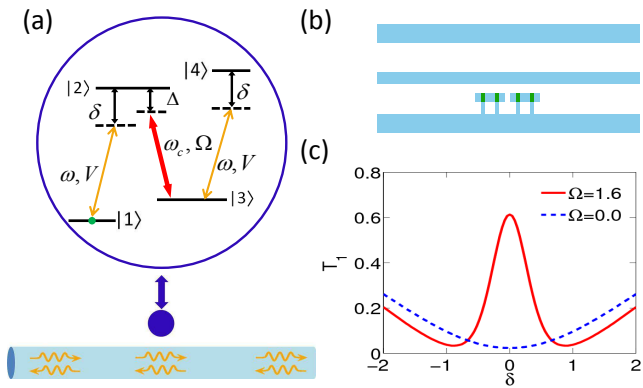}
\caption{Schematic diagram of the waveguide system and EIT. (a)  $N$-type four-level system interacting with the waveguide continuum. The transitions $|1\rangle\leftrightarrow|2\rangle$ and $|3\rangle\leftrightarrow|4\rangle$ are coupled to the waveguide modes with strength $V$, detuning $\delta$. The transition $|2\rangle\leftrightarrow|3\rangle$ is driven by a semiclassical control field with Rabi frequency $\Omega$ and detuning $\Delta$. Here, $\omega$ and $\omega_c$ are the frequencies of the incoming photons and the control field.
(b) Schematic of possible experimental setup based on superconducting charge qubits \cite{RebicPRL09}. Two charge 
qubits, each formed from two Josephson junctions [dark gray (green) regions], are coupled capacitively to each other and to a transmission line (1D waveguide). 
(c) Single-photon transmission probability as a function of incident photon detuning. EIT occurs when the control field is on ($\Omega=1.6$, solid); when it is off, the 4LS becomes a two-level system and EIT disappears. 
%Here, $P=6$.
Here, the effective Purcell factor is $P=6$. 
            %[Parameters: $\Delta=0$, $\omega_{21}=\omega_{43}$, $\sigma=0.2$, $\Gamma=6$, and $\Gamma_2 =1$.]
}
 \label{fig:structure}
 \end{figure} 

%\subsection*{\textcolor{blue}{\textbf{System and Hamiltonian}}}
Motivated by recent experimental advances \cite{AstafievSci10}, we consider an $N$-type 4LS \cite{MajerPRL05,RebicPRL09} coupled to a continuum of modes in a 1D waveguide. Figure 1 shows both a schematic and a possible realization using superconducting circuits \cite{RebicPRL09}. 
%The superconducting transmission line (horizontal bars) acts as a 1D waveguide, while the 4LS can be formed by placing two Cooper pair boxes (CPBs) together (Fig.~1b) \cite{RebicPRL09}. The capacitive coupling between the transmission line and the CPBs gives rise to an electric dipole interaction. 
The Hamiltonian of the system is \cite{ShenPRL07, ZhengPRA10}
\begin{eqnarray}
\lefteqn{H =\int\!\! dx\,(-i)\hbar c\Big[a_{R}^{\dagger}(x)\frac{d}{dx}a_{R}(x)
-a_{L}^{\dagger}(x)\frac{d}{dx}a_{L}(x)\Big]}& & \nonumber\\%+H_{\rm 4LS}  \nonumber \\[6pt]
 & +&\! \int\!\! dx\, \hbar V \delta(x)\big\{[a_{R}^{\dagger}(x)+a_{L}^{\dagger}(x)] 
 \big(|1\rangle \langle 2|+|3\rangle \langle 4|\big)+{\rm h.c.}\big\}, \nonumber \\
& +& \sum_{j=2}^{4} \hbar\Big(\epsilon_j-\frac{i\Gamma_j}{2}\Big)|j\rangle \langle j|
                  +\frac{\hbar \Omega}{2}\Big(|2\rangle \langle 3|+{\rm h.c.} \Big), 
\label{eq:Ham}
\end{eqnarray}
where $a_{L,R}^{\dagger}(x)$ are the creation operators for left,right-going photons at position $x$ and $c$ is the group velocity of photons. In the 4LS, the energy reference is the energy of state $|1\rangle$ (the ground state), and $\epsilon_2=\omega_{21}$, $\epsilon_3=\epsilon_2-\Delta$, and $\epsilon_4=\epsilon_3+\omega_{43}$, where $\omega_{21}$ and $\omega_{43}$ are the $|1\rangle\leftrightarrow|2\rangle$ and $|3\rangle\leftrightarrow|4\rangle$ transition frequencies, respectively. In the spirit of the quantum jump picture \cite{Carmichael93}, an imaginary term is included in the 4LS to model the spontaneous emission of the excited states at rate $\Gamma_j$ to modes other than the 1D waveguide continuum. We have assumed a linear dispersion and a frequency-independent coupling strength $V$ for the relevant frequency range \cite{ChangNatPhy07}. The decay rate into the waveguide modes is $\Gamma \equiv 2V^2/c$ from Fermi's golden rule. Below, we assume that level $|3\rangle$ is metastable ($\Gamma_3=0$) and levels $|2\rangle$ and $|4\rangle$ have the same loss rate ($\Gamma_2=\Gamma_4$).

A figure of merit to characterize the coupling strength is given by the effective Purcell factor, $P=\Gamma/\Gamma_2$. In an experiment with surface plasmons coupled to a quantum dot \cite{AkimovNat07}, $P=1.5$ was achieved. In more recent experiments, even larger Purcell factors, $P=3$ and $P\geq9$, were demonstrated with a superconducting transmission line \cite{AstafievSci10} and a GaAs photonic nanowire \cite{ClaudonNatPhoton10,BleusePRL11}, respectively. These recent dramatic experimental achievements suggest that the large Purcell-factor physics, which we now discuss, is presently within reach experimentally. 

%\subsection*{\textcolor{blue}{\textbf{Scattering eigenstates}}}

To study interaction effects during photon transmission, we obtain an exact solution of the scattering problem defined by Eq.\,(\ref{eq:Ham}). The scattering eigenstates are obtained by imposing an open boundary condition and requiring that the incident photon state be a free plane wave, an approach adopted previously to solve an interacting resonant-level model \cite{NishinoPRL09} and a two-level system problem \cite{ZhengPRA10}. 
%By generalizing the approach to a 4LS, we obtain one-photon, two-photon, and three-photon scattering eigenstates (see the Supplementary Information for details). 
For an incident photon from the left (with wave-vector $k$) and the 4LS initially in its ground state, the transmitted part of the single-photon eigenstate is
\begin{eqnarray}
\lefteqn{\phi_R(x) = t_ke^{ikx} } &&\nonumber \\
 & & t_k=\frac{[ck-\epsilon_{2}+\Delta+\frac{i\Gamma_{3}}{2}][ck-\epsilon_{2}+\frac{i\Gamma_{2}}{2}]-\frac{\Omega^{2}}{4}}{[ck-\epsilon_{2}+\Delta+\frac{i\Gamma_{3}}{2}][ck-\epsilon_{2}+\frac{i\Gamma_{2}}{2}+\frac{i\Gamma}{2}]-\frac{\Omega^{2}}{4}}, \quad
\label{eq:One-photon Transmission}
\end{eqnarray}
where $t_k$ is the transmission coefficient.
In the two-photon scattering eigenstate, the transmitted wave is
\begin{eqnarray}
\phi_{RR}(x_1,x_2)&=& t_{k_1}t_{k_2}(e^{ik_1x_1+ik_2x_2}+e^{ik_1x_2+ik_2x_1}) \nonumber \\
&+& b_1e^{-\gamma_1|x_2-x_1|}+b_2e^{-\gamma_2|x_2-x_1|},
\label{eq:Two-photon Transmission}
\end{eqnarray}
where $b_{1,2}$ and $\gamma_{1,2}$ ($>0$) are functions of system parameters (see the Supplementary Information). 
The first term of $\phi_{RR}$ corresponds to transmission of the two photons as independent (identical) particles with the momentum of each photon conserved individually. The second term is a two-body bound state---note the exponential decay in the relative coordinate $|x_2-x_1|$---with \emph{two} characteristic binding strengths $\gamma_{1}$ and  $\gamma_{2}$. Such a state 
results from %is a manifestation of 
the nonlinear interaction between photons mediated by the 4LS. Physically, it originates from the momentum non-conserved processes of each individual photon (with conservation of total momentum). A similar bound state has been found in a 1D waveguide coupled to a two-level system \cite{ShenPRL07,ZhengPRA10}, a $\Lambda$-type three-level system \cite{RoyPRL11}, as well as an open interacting resonant level model \cite{NishinoPRL09}.

%We emphasize that the existence of a continuum is essential for the emergence of the bound state: it is a unique many-body effect occurring in the waveguide system, but not in a single-mode cavity. Note also that a bound state is only possible when there are at least two photons interacting with the same 4LS: this leads to completely different transmission properties for a single photon and for two or more photons, which is the key to the appearance of a photon blockade.

%\subsection*{\textcolor{blue}{\textbf{EIT, photon blockade, and photon-induced tunneling}}}

We evaluate the transmission and reflection probabilities using the S-matrices constructed from the exact scattering eigenstates \cite{ZhengPRA10}.
Because any state containing a finite number of photons is, in practice, a wave packet, we consider a continuous mode input state \cite{LoudonQTL03}, whose spectrum is Gaussian with central frequency $\omega_0$ and width $\sigma$. Throughout this paper, we set the loss rate as the reference unit for all other quantities: $\Gamma_2=\Gamma_4=1$.  The Purcell factor becomes $P=\Gamma/\Gamma_2=\Gamma$. For all the numerical results shown, we take the detuning of the control field to be zero, $\Delta=0$, and choose $\sigma= \Gamma_2/5=0.2$. In addition, we assume that the transitions $|1\rangle\leftrightarrow|2\rangle$ and $|3\rangle\leftrightarrow|4\rangle$ are at the same frequency ($\omega_{21}=\omega_{43}$).

Figure 1(c) shows that the transmission probability $T_1$ of a single photon has a sharp peak as a function of its detuning $\delta \equiv \omega_0 - \omega_{21}$, demonstrating the familiar EIT phenomenon \cite{WitthautNJP10,RoyPRL11} produced by interference between two scattering pathways. The width of the EIT peak is $\sim\!\Omega$ for $P\gg1$ and is mainly determined by the control field strength $\Omega$. Here, we use $P=6$, a conservative value given the recent advances in experiments \cite{ClaudonNatPhoton10,AstafievSci10,BleusePRL11}. When the control field is turned off, the 4LS becomes a two-level system, which acts as a reflective mirror \cite{ChangNatPhy07}.

 \begin{figure}[t]
 \centering
 \includegraphics[width=0.32\textwidth]{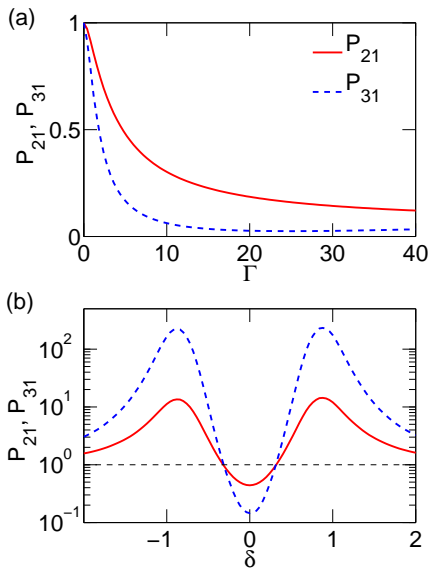}
 \caption{Photon blockade in transmission.
          (a) Photon blockade strength $P_{21}$ (solid) and $P_{31}$ (dashed) as a function of the decay rate $\Gamma$ into the waveguide modes (photon blockade is strongest when $P_{21}$ is smallest). The photons are on resonance with the 4LS ($\delta=0$) and $\Omega=1.6$. Strong coupling produces substantial photon blockade despite EIT being present in the single photon properties. 
          (b) $P_{21}$ and $P_{31}$ as a function of the incident photon detuning $\delta$ for $\Gamma=6$ and $\Omega=1.6$. There is clear photon blockade on resonance as well as photon-induced tunneling ($P_{21},P_{31}>1$) on the sides of the resonance.
         % [Parameters: $\Gamma=6$ ($P=6)$ and $\Omega=1.6$.]
          }
 \label{fig:blockade}
 \end{figure}

The EIT picture changes dramatically when there are two or more photons injected into the system. The 4LS mediates an effective photon-photon interaction, which in turn affects the multi-photon transmission. We define $T_2$ ($T_3$) to be the two- (three-) photon transmission probability of the two- (three-) photon scattering process. The strength of the photon blockade $P_{21}$ is given by the conditional probability $T_2/T_1$ for transmitting a second photon given that the first photon has already been transmitted, normalized by the single-photon transmission probability $T_1$: $P_{21}\equiv T_2/T_1^2$. For independent photons, there is no photon blockade and $P_{21}=1$. In the opposite limit of strong photon blockade, $P_{21}$ is suppressed towards zero. Similarly, we define $P_{31}\equiv T_3/T_1^3$ to quantify photon blockade in the three photon case. 

A pronounced photon blockade is shown in Fig.~2(a) in the strong coupling regime: the single photon EIT effect does \emph{not} carry over to the multi-photon case. For increasing coupling strength, both $P_{21}$ and $P_{31}$ approach zero. Such photon blockade regulates the flow of photons in an ordered manner, enabling coherent control over the information transfer process in our cavity-free scheme. 
%In addition, this makes possible the realization of all-optical switches and quantum logic gates without the need of a cavity. 
Taking $P=9$ achieved in Ref.\,\onlinecite{ClaudonNatPhoton10}, we obtain the values $P_{21}\sim30\%$ and $P_{31}\sim7\%$, showing that the effects predicted here are already within reach of experiments.

The photon blockade occurs despite being in the EIT regime: as shown in Fig.~2(b), both $P_{21}$ and $P_{31}$ are suppressed within the EIT window, whose width is set by the control field strength $\Omega$. However, away from the EIT window, $P_{21}$ and $P_{31}$ become larger than $1$, signaling a new regime of multi-photon transmission---photon-induced tunneling \cite{FaraonNatPhy08}. Previously, photon blockade and photon-induced tunneling have been observed in cavity-QED and circuit-QED systems \cite{BirnbaumNat05, FaraonNatPhy08,LangPRL11}, where the underlying mechanism is the anharmonicity of the spectrum caused by the atom-cavity coupling. We emphasize that such anharmonicity is absent in the present cavity-free scheme: the photon blockade and photon-induced tunneling here must be caused by a different mechanism.

 \begin{figure}[t]
 \centering
 \includegraphics[width=0.32\textwidth]{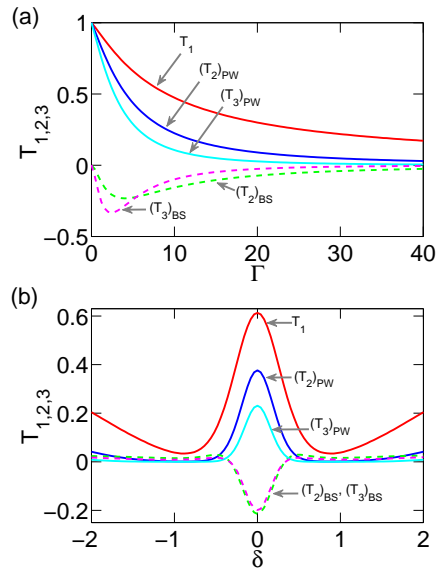}
 \caption{Many-body bound-state effect causing photon blockade.
        %  The two and three photon transmission probabilities are separated into the independent plane-wave part, labeled ``PW'', and the bound state and interference part, labeled ``BS'', and compared to the one photon transmission $T_1$. 
          (a) Transmission as a function of the decay rate $\Gamma$ into the waveguide modes for $\Omega=1.6$. The presence of the many-body bound states suppresses the multi-photon transmission. 
          (b) Transmission as a function of incident photon detuning $\delta$ for $\Gamma=6$ and $\Omega=1.6$. The bound states strikingly oppose multi-photon EIT; off-resonance, their constructive interference produces photon-induced tunneling. 
%       Same parameters as for Fig.\,2.
%   [Parameters: $\Gamma=6$ ($P=6)$ and $\Omega=1.6$.] 
                    }
 \label{fig:boundstate}
 \end{figure}

To understand the origin of these phenomena, we separate the two-photon transmission probability $T_2$ into two parts (Fig.~3): $(T_2)_{\rm PW}=T_1^2$ is the contribution from independent transmission and $(T_2)_{\rm BS}$ is the contribution from both the bound-state term in Eq.\,(\ref{eq:Two-photon Transmission}) and the interference between the plane-wave and bound-state terms. $(T_2)_{\rm BS}$ is the result of the many-body interactions in the waveguide and is absent for cavities. Similarly, $T_3$ can be separated into $(T_3)_{\rm PW}=T_1^3$ and $(T_3)_{\rm BS}$. Figure 3(a) shows that, when the photons are on resonance, $(T_{2,3})_{\rm BS}$ is always negative, suppressing the overall transmission. The cause of the observed photon blockade is, thus, the destructive interference between the two transmission pathways: passing by the 4LS as independent particles or as a composite particle in the form of bound states. Within the EIT window, this conclusion always holds (see Fig.\,3(b))---both two- and three-photon transmission are strongly suppressed by the many-body bound-state effect. In contrast, away from the EIT window, $(T_{2,3})_{\rm BS}$ changes sign and becomes positive (see the Supplementary Information for a detailed analysis of the interference). Destructive interference changes to constructive interference, producing photon-induced tunneling.

 \begin{figure}[t]
 \centering
 \includegraphics[width=0.45\textwidth]{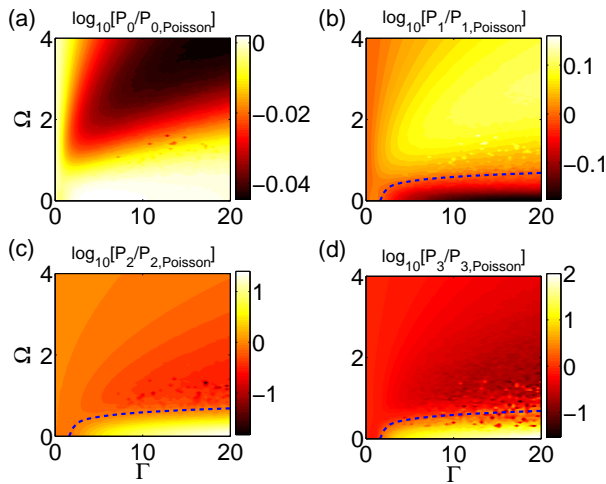}
 \caption{Non-classical photon states. Photon number statistics quantified by $\text{log}_{10}(P_n/P_{n,\text{Poisson}})$, where the probability of $n$ photons in the output state is $P_n$ and $P_{n,\text{Poisson}}$ is for a coherent state with the same mean photon number. Panels (a)-(d) show results for $n=0,1,2,$ and $3$, respectively, as a function of the control field strength $\Omega$ and the decay rate $\Gamma$. The dashed line is a guide to the eye indicating equal probabilities, $P_n/P_{n,\text{Poisson}}=1$. Note the large range of parameters for which the single photon probability is enhanced while the multi-photon content is suppressed---an improved single photon source.   
 }
 \label{fig:nonclassical}
 \end{figure}

%\subsection*{\textcolor{blue}{\textbf{Single-photon source}}}
%The existence of the strong nonlinear quantum optical phenomena discussed above---EIT, photon blockade, and photon-induced tunneling---in a cavity-free setting opens up new possibilities for coherent control over light. 
As an application, we now show that the 4LS can generate non-classical photon states.
We assume that the 4LS is in its ground state initially. We consider an incident continuous-mode coherent state of mean photon number $\bar{n}=1.0$, on resonance with the 4LS. The photon number statistics in the transmitted field can be obtained using the S-matrix method \cite{ZhengPRA10} and is presented in Fig.~4 by taking the ratio of the photon number distribution of the output state $P_n$ ($n=0,1,2,3$) to that of a coherent state $P_{n,\text{Poisson}}$ having the same mean photon number. It is remarkable that, in most of the parameter space, we have $P_1>P_{1,\text{Poisson}}$ while $P_2<P_{2,\text{Poisson}}$ and $P_3<P_{3,\text{Poisson}}$. This gives rise to a sub-Poissonian single-photon source: for example, for $\Gamma=9$ ($P=9$) and $\Omega=1.6$, we have $P_0=64\%$ and $P_1=34\%$ with the multi-photon probability $P_{n\geq2}$ less than $3\%$, in comparison with $P_{n\geq2}=26\%$ in the input. This single-photon source comes about because, under EIT conditions, a single photon passes through the system with high probability, while multi-photon states experience photon blockade caused by the bound-state effect. A systematic way of improving the quality of the single-photon source is to let a coherent state with a large mean photon number pass through multiple 4LS devices in series with Faraday isolators inserted between each stage.

In summary, we present a cavity-free scheme to realize a variety of nonlinear quantum optical phenomena---including EIT, photon blockade and photon-induced tunneling---in a 1D waveguide. Photon blockade and photon-induced tunneling have a distinctly different origin here compared to the cavity case: a many-body bound-state effect. Furthermore, we outline how to use EIT and photon blockade in this system to produce a single-photon source on demand. 
On the one hand, the demonstrated ability to control the flow of light quanta using EIT and photon blockade is a critical step towards the realization of an open quantum network. 
On the other hand, the strong photon-photon interaction mediated by the 4LS provides a new candidate system to study strongly correlated 1D systems, one complementary to condensed-matter systems.

%\subsection*{\textcolor{blue}{\textbf{Acknowledgements}}}
%\bigskip\textcolor{blue}{\textbf{Acknowledgements:}}
The work of HUB and HZ was supported by the U.S.\,Office of Naval Research.

\vspace*{-0.25in}
%\bibliographystyle{unsrt}
%\bibliography{4LS_PRL}

\newpage
\widetext

\begin{center}
\large\bf Supplementary Material for ``Cavity-free Photon
Blockade Induced by Many-body Bound States''
\end{center}

\global\long\def\theequation{S\arabic{equation}}

\subsection*{Hamiltonian}

The Hamiltonian describing the interaction between a 1D continuum
and a 4LS at the spatial origin is given by \cite{SM_ChangNatPhy07,SM_ShenPRA09}, 

\begin{eqnarray}
H & = & \int dk\,\hbar\omega_{k}a^{\dagger}(k)a(k)+H_{4LS}+\int dk\,\hbar g\Big\{ a^{\dagger}(k)\big[S_{12}+S_{34}\big]+h.c.\Big\},\nonumber \\
H_{4LS} & = & \sum_{j=2}^{4}\hbar\left(\epsilon_{j}-\frac{i\Gamma_{j}}{2}\right)|j\rangle\langle j|+\frac{\hbar\Omega}{2}\left(|2\rangle\langle3|+h.c.\right)\label{eq:Hamil}\end{eqnarray}

\noindent where $g$ is the coupling strength, $S_{12}=|1\rangle\langle2|$
and $S_{34}=|3\rangle\langle4|$. Linearizing the dispersion around
the transition energy of the 4LS and treating the left-going and right-going
photons as separate fields, we obtain a two-mode Hamiltonian. Transformation
to the real space field operators $a_{L(R)}(x)=(1/\sqrt{2\pi})\int dke^{ikx}a_{L(R)}(k)$
yields the real-space Hamiltonian

\begin{equation}
H=\int dx\hbar c[a_{R}^{\dagger}(x)(-i\frac{d}{dx})a_{R}(x)-a_{L}^{\dagger}(x)(-i\frac{d}{dx})a_{L}(x)]+H_{4LS}+\int dx\hbar V\Big\{[a_{R}^{\dagger}(x)+a_{L}^{\dagger}(x)]\big[S_{12}+S_{34}\big]+h.c.\Big\},\label{eq:Hamil2}\end{equation}

\noindent where $V=\sqrt{2\pi}g$. Finally, the coupling between the
4LS and general modes other than the waveguide can be incorporated
in the same manner as the waveguide-4LS interaction. Such interaction
with general reservoir modes, such as vacuum modes, leads to intrinsic
dissipation. This term can be treated by the Wigner-Weisskopf approximation
\cite{SM_Meystre} and so replaced by a non-Hermitian damping term in
$H_{4LS}$ \cite{SM_ShenPRA09,SM_Carmichael93}, thus giving rise to the
Hamiltonian in Eq. (1) of the main text.

\subsection*{Scattering Eigenstates}

A key insight to solve for the scattering eigenstates is gained by
performing the operator transformation

\begin{equation}
a_{e/o}^{\dagger}(x)\equiv\big[a_{R}^{\dagger}(x)\pm a_{L}^{\dagger}(-x)\big]/\sqrt{2}.\label{eq:}\end{equation}

\noindent The Hamiltonian is then decomposed into even and odd modes:
$H=H_{e}+H_{o}$ with

\begin{eqnarray}
H_{e} & = & \int dx(-i)\hbar c\hat{a}_{e}^{\dagger}(x)\frac{d}{dx}\hat{a}_{e}(x)+H_{4LS}\nonumber \\
 &  & +\int dx\,\hbar\sqrt{2}V\,\delta(x)\big[\hat{a}_{e}^{\dagger}(x)(S_{12}+S_{34})+h.c.\big],\label{eq:Heo}\\
H_{o} & = & \int dx(-i)\hbar c\hat{a}_{o}^{\dagger}(x)\frac{d}{dx}\hat{a}_{o}(x).\nonumber \end{eqnarray}

\noindent Because $[H,\hat{n}_{e}+\hat{n}_{{\rm 4LS}}]=[H,\hat{n}_{o}]=0$
for the number operators $\hat{n}_{e/o}\equiv\int dx\hat{a}_{e/o}^{\dagger}(x)\hat{a}_{e/o}(x)$
and $\hat{n}_{{\rm 4LS}}\equiv\sum_{j=2,3}|j\rangle\langle j|+2|4\rangle\langle4|$,
the total number of excitations in both the even and odd spaces are
separately conserved. We will now focus on finding the non-trivial
even-mode solution and then transform back to the left/right representation.
A general $n$-excitation state ($n=n_{e}+n_{{\rm 4LS}}$) in the
even space is given by \begin{eqnarray}
|\Psi_{n}\left\rangle \right. & = & \bigg(\int dx^{n}\, g^{(n)}(x)\,\hat{a}_{e}^{\dagger}(x_{1})\cdots\hat{a}_{e}^{\dagger}(x_{n})\nonumber \\[6pt]
 & + & \int dx^{n-1}\,\big[f_{2}^{(n)}(x)S_{12}^{+}+f_{3}^{(n)}(x)S_{13}^{+}\big]\ \hat{a}_{e}^{\dagger}(x_{1})\cdots\hat{a}_{e}^{\dagger}(x_{n-1}),\nonumber \\
 & + & \int dx^{n-2}\, f_{4}^{(n)}(x)S_{14}^{+}\hat{a}_{e}^{\dagger}(x_{1})\cdots\hat{a}_{e}^{\dagger}(x_{n-2})\bigg)|0,1\rangle\end{eqnarray}
 where $|0,1\rangle$ is the zero photon state with the 4LS in the
ground state $|1\rangle$. 

The scattering eigenstates are constructed by imposing the boundary
condition that, in the incident region $x_{1},\cdots,x_{n}<0$, $g^{(n)}(x_{1},\cdots,x_{n})$
is a free plane wave \cite{SM_NishinoPRL09,SM_ZhengPRA10}. The one-photon
scattering eigenstate with eigenenergy $E=ck$ is given by \begin{eqnarray}
g^{(1)}(x) & = & g_{k}(x)=h_{k}(x)[\theta(-x)+\bar{t}_{k}\theta(x)],\,\,\,\,\,\,\,\,\,\,\,\,\,\,\,\, h_{k}(x)\equiv\frac{e^{ikx}}{\sqrt{2\pi}},\,\,\,\nonumber \\[6pt]
f_{2}^{(1)} & = & \frac{ic}{\sqrt{2\pi\Gamma}}(\bar{t}_{k}-1),\label{eq:OneSol}\\[6pt]
f_{3}^{(1)} & = & \frac{ic\Omega}{\sqrt{8\pi\Gamma}}\frac{(\bar{t}_{k}-1)}{ck+\Delta+i\Gamma_{3}/2-\epsilon_{2}},\nonumber \\[6pt]
\bar{t}_{k} & = & \frac{[ck-\epsilon_{2}+\Delta+i\Gamma_{3}/2][ck-\epsilon_{2}+i\Gamma_{2}/2-i\Gamma/2]-\Omega^{2}/4}{[ck-\epsilon_{2}+\Delta+i\Gamma_{3}/2][ck-\epsilon_{2}+i\Gamma_{2}/2+i\Gamma/2]-\Omega^{2}/4},\nonumber \end{eqnarray}
 where $\theta(x)$ is the step function and $\Gamma=2V^{2}/c$ is
the spontaneous emission rate to the 1D continuum. Notice that the
amplitude $f_{4}^{(1)}$ vanishes because it takes two excitation
quanta to excite level $|4\rangle$. Therefore, for one-photon scattering,
the 4LS behaves as a three-level system and Eq. (\ref{eq:OneSol})
coincides with previous studies \cite{SM_WitthautNJP10,SM_RoyPRL11}. Transforming
from the even/odd back to the left/right representation gives rise
to the wavefunction of one photon transmission in the single-photon
scattering, shown in Eq. (2) in the main text, where $t_{k}=(\bar{t}_{k}+1)/2$.

For two-photon scattering, starting from a free plane wave in the
region $x_{1},x_{2}<0$, we use the Schrodinger equation implied by
$H_{e}$ in Eq. (\ref{eq:Heo}) to find the wave function first in
the region $x_{1}<0<x_{2}$ and then for $0<x_{1},x_{2}$. The method
is explained in detail in the appendix of Ref. \cite{SM_ZhengPRA10}
for scattering from a two level system. We obtain the two-photon scattering
eigenstate with eigenenergy $E=ck_{1}+ck_{2}$ \begin{eqnarray}
g^{(2)}(x_{1},x_{2}) & = & \frac{1}{2}\sum_{Q}\left[g_{k_{1}}(x_{Q_{1}})g_{k_{2}}(x_{Q_{2}})+B(x_{Q_{1}},x_{Q_{2}})\right],\nonumber \\
f_{2}^{(2)}(x) & = & \sum_{Q}\frac{ic(\bar{t}_{k_{Q_{1}}}-1)g_{k_{Q_{2}}}(x)}{\sqrt{2\pi\Gamma}}+\frac{ice^{iEx}}{\sqrt{\Gamma}}[C_{1}e^{\lambda_{1}x}+C_{2}e^{\lambda_{2}x}]\theta(x),\nonumber \\
f_{3}^{(3)}(x) & = & \frac{c\Omega\sqrt{\Gamma}}{\sqrt{8\pi}}\sum_{Q}\frac{g_{k_{Q_{1}}}(x)}{\rho_{k_{Q_{2}}}}-\frac{2ce^{iEx}}{\sqrt{\Gamma}\Omega}[\kappa_{1}C_{1}e^{\lambda_{1}x}+\kappa_{2}C_{2}e^{\lambda_{2}x}]\theta(x),\nonumber \\
f_{4}^{(2)} & = & \sum_{Q}\frac{ic\Omega(\bar{t}_{k_{Q_{1}}}-1)}{4\pi\rho_{k_{Q_{2}}}}-\frac{2ic\beta}{\Gamma\Omega},\label{eq:TwoSol}\\
B(x_{Q_{1}},x_{Q_{2}}) & = & e^{iEx_{Q_{2}}}[C_{1}e^{\lambda_{1}|x_{2}-x_{1}|}+C_{2}e^{\lambda_{2}|x_{2}-x_{1}|}]\theta(x_{Q_{2}}-x_{Q_{1}})\theta(x_{Q_{1}}),\nonumber \end{eqnarray}
 where $Q=(Q_{1},Q_{2})$ is a permutation of $(1,2)$, and \begin{eqnarray}
c\lambda_{1} & = & -\frac{\Gamma+\Gamma_{2}+\Gamma_{3}}{4}+\xi+i(\frac{\Delta}{2}-\epsilon_{2}+\eta),\,\,\,\,\,\,\,\,\,\,\,\,\, c\lambda_{2}=-\frac{\Gamma+\Gamma_{2}+\Gamma_{3}}{4}-\xi+i(\frac{\Delta}{2}-\epsilon_{2}-\eta)\nonumber \\
C_{1} & = & \frac{\beta-\alpha c\lambda_{2}}{\lambda_{1}-\lambda_{2}},\,\,\,\,\,\,\,\,\,\,\,\,\, C_{2}=-\frac{\beta-\alpha c\lambda_{1}}{\lambda_{1}-\lambda_{2}},\nonumber \\
\rho_{k} & = & \left(ck-\epsilon_{2}+\Delta+\frac{i\Gamma_{3}}{2}\right)\left(ck-\epsilon_{2}+\frac{i\Gamma_{2}+i\Gamma}{2}\right)-\frac{\Omega^{2}}{4},\,\,\,\,\,\,\,\,\,\,\,\,\,\chi=\Delta^{2}+\Omega^{2}-\Gamma^{\prime2}\nonumber \\
\kappa_{1} & = & \frac{\Gamma^{\prime}}{2}+\xi+i(\frac{\Delta}{2}+\eta),\,\,\,\,\,\,\,\,\,\,\,\,\,\,\,\,\,\,\,\,\,\,\,\,\,\,\,\kappa_{2}=\frac{\Gamma^{\prime}}{2}-\xi+i(\frac{\Delta}{2}-\eta),\nonumber \\
\xi & = & \frac{\sqrt{2}}{4}\sqrt{\sqrt{\chi^{2}+4\Delta^{2}\Gamma^{\prime2}}-\chi},\,\,\,\,\,\,\,\,\,\,\,\,\,\eta=\frac{\sqrt{2}}{4}\sqrt{\sqrt{\chi^{2}+4\Delta^{2}\Gamma^{\prime2}}+\chi}\nonumber \\
\nonumber \\\alpha & = & -\frac{(\bar{t}_{k_{1}}-1)(\bar{t}_{k_{2}}-1)}{\pi},\,\,\,\,\,\,\,\,\,\,\,\,\,\,\,\,\,\,\,\,\,\,\,\,\,\,\,\beta=\frac{-\Gamma\Omega^{2}}{8\pi}\left[\frac{\nu-\bar{t}_{k_{1}}}{\rho_{k_{2}}}+\frac{\nu-\bar{t}_{k_{2}}}{\rho_{k_{1}}}\right],\nonumber \\
\Gamma^{\prime} & = & \frac{\Gamma+\Gamma_{2}-\Gamma_{3}}{2},\,\,\,\,\,\,\,\,\,\,\,\,\,\,\,\,\,\,\,\,\,\,\,\,\,\,\,\,\,\,\,\,\,\,\,\,\,\,\,\,\,\,\,\,\,\,\nu=\frac{\epsilon_{4}-E-(i\Gamma_{4}-i\Gamma)/2}{\epsilon_{4}-E-(i\Gamma_{4}+i\Gamma)/2}.\end{eqnarray}

\noindent Again, performing an even/odd to left/right transformation
\cite{SM_ZhengPRA10}, we obtain the wavefunction of two photon transmission
in two-photon scattering, shown in Eq. (3) in the main text. There,
$\gamma_{1/2}=-\text{Re}[\lambda_{1/2}]$ and $b_{1/2}\propto C_{1/2}$.
In the case that $\Delta=\Gamma_{3}=0$, $\xi=|\chi|\theta(-\chi)/2$
and $\eta=|\chi|\theta(\chi)/2$ with $\chi=\Omega^{2}-\left(\Gamma+\Gamma_{2}\right)^{2}/4$,
where again $\theta(x)$ is the step function.

From the scattering eigenstates, scattering matrices can be constructed
using the Lippmann-Schwinger formalism \cite{SM_Sakurai}. The output
states are then obtained by applying the scattering matrices on the
incident photon states (single-, two-, and three-photon number states
as well as coherent states). By performing measurement operations
on the output states, we obtain the transmission and reflection probabilities
and photon number statistics. A general procedure is outlined in Ref.
\cite{SM_ZhengPRA10}.

\begin{figure}
\begin{centering}
\includegraphics[scale=0.5]{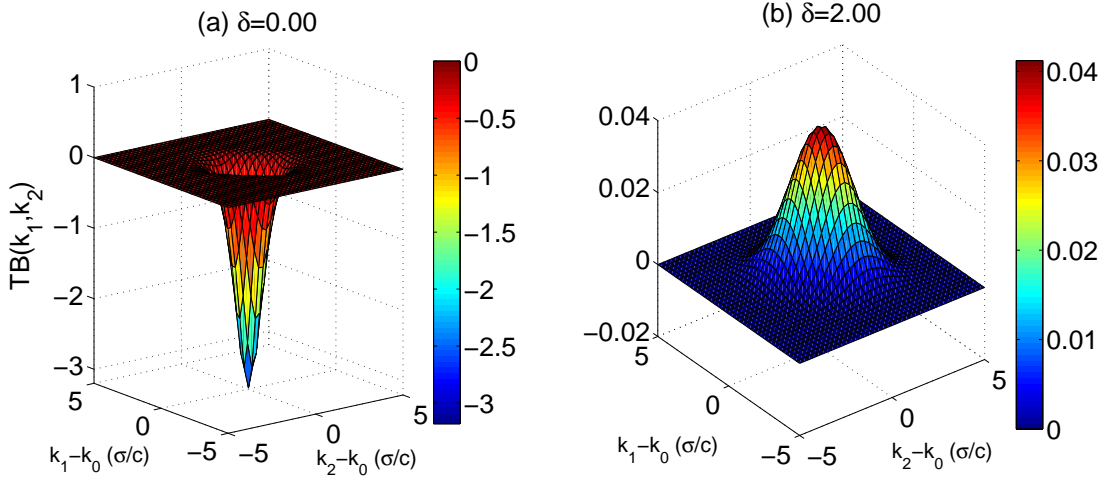}
\par\end{centering}

\caption{The integrand function $TB(k_{1},k_{2})$ of the interference term
as a function of $k_{1}$and $k_{2}$. (a) $\delta=0.0$. (b) $\delta=2.0$.}
\label{fig:phase analysis}
\end{figure}

\subsection*{Phase Analysis of Photon Blockade and Photon-Induced Tunneling}

In Fig. 3 of the main text, we show that the observed photon blockade
and photon-induced tunneling orignate from destructive and constructive
interference effects, respectively. In this section, we present a
detailed analysis of the relative phase between the plane-wave and
bound-state terms. The two-photon transmission probability $T_{2}$
for the two-photon scattering is given by

\begin{equation}
T_{2}=\int dk_{1}dk_{2}|T(k_{1},k_{2})+B(k_{1},k_{2})|^{2},\label{eq:T2}\end{equation}

\noindent where $k_{1}$ and $k_{2}$ are the momenta of the two photons
in the wavepacket. The plane-wave term $T(k_{1},k_{2})$ and the bound-state
term $B(k_{1},k_{2})$ take the form

\begin{eqnarray}
T(k_{1},k_{2}) & = & \alpha(k_{1})\alpha(k_{2})t_{k_{1}}t_{k_{2}}\nonumber \\
B(k_{1},k_{2}) & = & \sum_{j=1,2}\frac{\sqrt{\Gamma}}{8}\left(\frac{1}{k_{1}-i\gamma_{j}}+\frac{1}{k_{2}-i\gamma_{j}}\right)\int dk^{\prime}\alpha(k^{\prime})\alpha(k_{1}+k_{2}-k^{\prime})C_{j}(k^{\prime},k_{1}+k_{2}-k^{\prime}),\label{eq:TB}\end{eqnarray}

\noindent where $\alpha(k)$ is the amplitude of a Gaussian wavepacket
with central frequency $\omega_{0}$ and width $\sigma$. In momentum
space, the central momentum is $k_{0}=\omega_{0}/c$ and the width
is $\sigma/c$, where $c$ is the group velocity of photons in the
waveguide. By defining the phase difference $\theta(k_{1},k_{2})$
between $T(k_{1},k_{2})$ and $B(k_{1},k_{2})$, $T_{2}$ can be written
as

\begin{equation}
T_{2}=\int dk_{1}dk_{2}|T(k_{1},k_{2})|^{2}+\int dk_{1}dk_{2}2|T(k_{1},k_{2})B(k_{1},k_{2})|\text{cos}\theta(k_{1},k_{2})+\int dk_{1}dk_{2}|B(k_{1},k_{2})|^{2},\label{eq:T2_Separated}\end{equation}

\noindent where the first term is the plane-wave term, the second
term is the interference between the plane-wave and bound-state terms,
and the third term is the contribution from the bound-state term.
In the main text, the first term is denoted $(T_{2})_{PW}$ and the
second and third terms together are called $(T_{2})_{BS}.$ Denote
the intergrand function of the interference term as $TB(k_{1},k_{2})$,

\begin{equation}
TB(k_{1},k_{2})=2|T(k_{1},k_{2})B(k_{1},k_{2})|\text{cos}\theta(k_{1},k_{2})\label{eq:TBk1k2}\end{equation}

\noindent The phase $\theta(k_{1},k_{2})$, or specifically, the sign
of $\text{cos}\theta(k_{1},k_{2})$, determines whether the interference
is constructive or destructive. 

We numerically evaluate $TB(k_{1},k_{2})$ in two cases: $\delta=\omega_{0}-\omega_{21}=0.0$
and $\delta=\omega_{0}-\omega_{21}=2.0$. Here, the unit of detuning
$\delta$ is set by the loss rate $\Gamma_{2}$. All the other system
parameters are the same as in Fig. 3b in the main text. Fig.~\ref{fig:phase analysis} shows
$TB(k_{1},k_{2})$ as a function of $k_{1}-k_{0}$ and $k_{2}-k_{0}$.
As expected for a Gaussian packet, the value of $TB(k_{1},k_{2})$
is centered at $k_{1}=k_{2}=k_{0}$ in both cases. However, the sign
of the peaks in the two cases differs. For $\delta=0.0$, a negative
peak indicates destructive interference, giving rise to photon blockade
when the incident photons are on resonane with the 4LS. For $\delta=2.0$,
a positive peak indicates constructive interference, producing photon-induced
tunneling when the photons are far off resonance.

%\bibliography{WQED}

\end{document}